\documentclass{PoS}

\usepackage{amsmath}

\newcommand{\be}{\begin{equation}}
\newcommand{\ee}{\end{equation}}
\newcommand{\bea}{\begin{eqnarray}}
\newcommand{\eea}{\end{eqnarray}}
\newcommand{\bt}{\begin{tabbing}}
\newcommand{\et}{\end{tabbing}}
\newcommand{\bi}{\begin{itemize}}
\newcommand{\ei}{\end{itemize}}
\newcommand{\ben}{\begin{enumerate}}
\newcommand{\een}{\end{enumerate}}
\newcommand{\nn}{\nonumber}

\newcommand{\calO}{{\mathcal O}}
\newcommand{\Dt}{\Delta t}
\newcommand{\Dtp}{\Delta t^\prime}
\newcommand{\bfp}{{\bf p}}
\newcommand{\bfpp}{{\bf p^\prime}}
\newcommand{\bfx}{{\bf x}}
\newcommand{\bfxp}{{\bf x^\prime}}
\newcommand{\bfxpp}{{\bf x^{\prime\prime}}}

\newcommand{\emFpip}{F_{V}^{\pi^+}}
\newcommand{\sFpi}{F_{S}^{\pi}}
\newcommand{\emFkp}{F_{V}^{K^+}}
\newcommand{\emFkn}{F_{V}^{K^0}}

\newcommand{\cradpip}{\langle r^2 \rangle_{V}^{\pi^+}}
\newcommand{\sradpi}{\langle r^2 \rangle_{S}^{\pi}}
\newcommand{\cradkn}{\langle r^2 \rangle_{V}^{K^0}}
\newcommand{\cradkp}{\langle r^2 \rangle_{V}^{K^+}}

\title{
   \begin{picture}(0,0)(0,0)%
   \put(355,75){\makebox(0,0)[l]{\textnormal{\normalsize KEK-CP-245}}}%
   \end{picture}%
   Light meson form factors in $N_f=2+1$ QCD with dynamical overlap quarks
}

\ShortTitle{
   Light meson form factors in $N_f=2+1$ QCD with dynamical overlap quarks
}

\author{ 
   JLQCD Collaboration: 
   \speaker{T.~Kaneko}$^{a,b}$\thanks{E-mail: takashi.kaneko@kek.jp}, 
   S.~Aoki$^{c,d}$, 
   G.~Cossu$^a$, 
   H.~Fukaya$^e$, 
   S.~Hashimoto$^{a,b}$, 
   J.~Noaki$^{a}$
   and
   T.~Onogi$^e$
   \\
   \\
   \\
   \llap{$^a$}
   High Energy Accelerator Research Organization (KEK),
   Tsukuba 305-0801, Japan 
   \\
   \llap{$^b$}
   School of High Energy Accelerator Science,
   The Graduate University for Advanced Studies (Sokendai),
   Tsukuba 305-0801, Japan
   \\ 
   \llap{$^c$}
   Graduate School of Pure and Applied Sciences, 
   University of Tsukuba, Tsukuba 305-8571, Japan
   \\ 
   \llap{$^d$}
   Center for Computational Sciences, University of Tsukuba, 
   Tsukuba 305-8577, Japan
   \\
   \llap{$^e$}
   Department of Physics, Osaka University, 
   Toyonaka, Osaka 560-0043 Japan
}

\abstract{
We report on our calculation of pion and kaon form factors in three-flavor 
QCD using the overlap quark action.
Gauge ensembles are generated on a $16^3 \times 48$ lattice 
at a lattice spacing of 0.11 fm with pion masses down to 310~MeV.
Connected and disconnected meson correaltors are 
calculated using the all-to-all quark propagator. 
We present our preliminary analysis on the chiral behavior of 
the electromagnetic and scalar form factors as well as 
a comparison of the shape of the $K \to \pi$ form factors 
with experiment.
}

\FullConference{
The XXVIII International Symposium on Lattice Field Theory, Lattice2010\\
June 14-19, 2010\\
Villasimius, Italy
}

\begin{document}

% page limit: 7pages
% 1 : title
% 1 : intro 
% 1 : em
%       eff + q^2
%       chiral fit (pi)
%       q^2 
%       chiral fit (K)
% 1 : scalar
%       q^2 + chiral fit 
% 1 : weak 
%       xi 
%       f_0 fit + shape of f_+
% 1 concl. + ref 

%// introduction ==============================================================

\section{Introduction}
\vspace{-2mm}

One of the major goals of lattice QCD is a precise determination
of hadron form factors. 
% which are important inputs for 
% phenomenological studies of hadron processes.
%
Electromagnetic (EM) and scalar form factors
of pions and kaons are fundamental observables in hadron physics.
A detailed comparison between their chiral behavior on the lattice
and that in chiral perturbation theory (ChPT) 
may provide a good testing ground for these theoretical tools
as well as a determination of unknown parameters in ChPT,
namely the low-energy constants (LECs).
Reliable calculation of the form factors of semileptonic weak decays, 
such as the $K \! \to \! \pi l \nu$ decays, 
is important in the search for new physics
through a precise determination of CKM matrix elements.

In this article, 
we report on our calculation of these light meson form factors.
Chiral symmetry is exactly preserved in our simulations,  
that enables us to directly compare our results with ChPT.
We use the all-to-all quark propagator in order to accurately 
calculate connected and disconnected correlators of pions and kaons.

%// Simulations ===============================================================

\vspace{-1mm}
\section{Numerical simulations} 
\vspace{-2mm}

%// configuration generation 

%// gauge configuration

Our gauge ensembles of $N_f\!=\!2\!+\!1$ QCD are generated 
using the Iwasaki gauge and the overlap quark actions
with a topology fixing term \cite{exW+extmW:JLQCD},
which remarkably reduces the computational cost. 
While we only explore the trivial topological sector at the moment, 
the effect of fixing topology on the pion form factors 
turned out to be under control, typically a few \%, 
in our previous study in two-flavor QCD \cite{PFF:Nf2:RG+Ovr:JLQCD}.
On a $N_s^3 \! \times \! N_t \!=\! 16^3 \! \times \! 48$ lattice,
we simulate a single lattice spacing $a\!=\!0.112(1)$~fm
leaving a quantitative estimate of discretization errors for future studies.
Four values of degenerate up and down quark masses
$m_{ud}\!=\!0.015$, 0.025, 0.035 and 0.050
are taken to explore a range of the pion mass 
$310 \! \lesssim \! M_\pi[\mbox{GeV}] \! \lesssim 560$.
Measurements of the form factors are carried out 
with the periodic boundary condition
at a single value of the strange quark mass $m_s\!=\!0.080$, 
which is very close to the physical value 0.081 fixed from $M_K$.
We have accumulated 2,500 HMC trajectories 
at each combination of $m_{ud}$ and $m_s$.
% We note that our simulations on a larger volume $24^3 \! \times \! 48$
% with twisted boundary conditions \cite{TBC} are currently underway.
% In this article, we present preliminary analysis of currently available data.

%// measurement w/ a2a

We calculate meson correlators 
using the all-to-all quark propagator \cite{A2A}.
Let us consider an expansion of the propagator 
using the eigenmodes $(\lambda_k,u_k)$ ($k\!=\!1,\ldots,12 N_s^3 N_t)$ 
of the overlap-Dirac operator,
namely 
$D^{-1}(x,y) = \sum_k \lambda_k^{-1}u_k(x)u_k^\dagger(y)$.
It is expected that 
low-lying modes dominantly contribute to 
low-energy observables, such as the form factors.
We exactly evaluate this important contribution
by using 160 eigenmodes for each configuration.
Remaining contribution from the higher modes 
is estimated stochastically by using the noise method.
We refer readers to Ref.~\cite{PFF:Nf2:RG+Ovr:JLQCD} 
for further details on our measurement method using the all-to-all propagator.

We calculate three- and two-point functions 
\bea
   C^{P \calO P^\prime}_{\phi \phi^\prime}(\Dt,\Dtp,\bfp,\bfpp)
   & = & 
   \frac{1}{N_t N_s^3}\sum_{\bfx,t}
   \sum_{\bfxp,\bfxpp}
   \langle 
      P^\prime_{\phi^\prime}(\bfxpp,t+\Dt+\Dtp) \calO(\bfxp,t+\Dt) P_\phi^\dagger(\bfx,t)
   \rangle
   \nn \\
   && \hspace{60mm}
   \times e^{-i\bfpp(\bfxpp-\bfxp)-i\bfp(\bfxp-\bfx)},
   \label{eqn:sim:a2a:corr_3pt}
   \\
   C^{P P^\prime}_{\phi \phi^\prime}(\Dt,\bfp)
   & = & 
   \frac{1}{N_t N_s^3}\sum_{\bfx,t}
   \sum_{\bfxp}
   \langle 
      P^\prime_{\phi^\prime}(\bfxp,t+\Dt) P_\phi^\dagger(\bfx,t)
   \rangle
   e^{-i\bfp(\bfxp-\bfx)},
   \label{eqn:sim:a2a:corr_2pt}
\eea  
where 
$P_\phi^\dagger$ ($P^{\prime \dagger}_{\phi^\prime}$) represents 
an interpolating operator for the initial (final) meson 
with a smearing function $\phi$ ($\phi^\prime$),
and $\calO$ is 
either the EM current ($J_\mu$), weak current ($V_\mu$), 
or scalar operator ($S$).
Using the all-to-all propagator,
we can accurately calculate these correlators
by taking the average over the location of the meson source, 
namely $(\bfx,t)$ in Eqs.~(\ref{eqn:sim:a2a:corr_3pt}) 
and (\ref{eqn:sim:a2a:corr_2pt}).
In addition, 
the correlators 
with different choices of $\phi^{(\prime)}$ and $\bfp^{(\prime)}$ 
can be calculated with small additional costs.
In this study, 
we take two choices of $\phi^{(\prime)}$, 
namely local $\phi_l(r)\!=\!\delta_{r,0}$ 
and exponential functions $\phi_s(r)\!=\!e^{-0.4r}$,
and 27 choices of $\bfp^{(\prime)}$ with $|\bfp^{(\prime)}|\!\leq\!\sqrt{3}$,
which cover a region of the momentum transfer 
$-2.0 \! \lesssim \! q^2[\mbox{GeV}^2] \! \lesssim 0$.
Note that meson momenta $p^{(\prime)}$ are shown in units of $2\pi/(N_s a)$
in this article.

%// electromagnetic form factor ===============================================

\vspace{-1mm}
\section{Electromagnetic form factors}
\vspace{-2mm}

\begin{figure}[t]
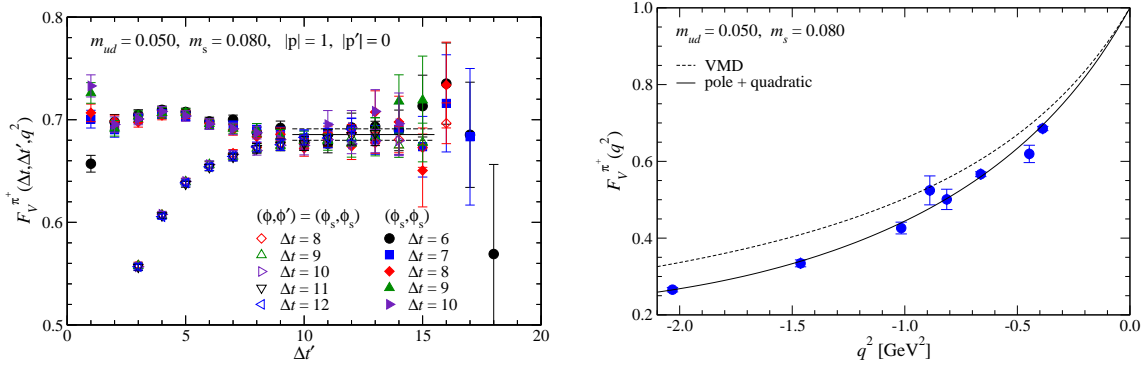

\begin{center}
\includegraphics[angle=0,width=0.48\linewidth,clip]%
                {pff_vs_dtsnk_mom0100.lat10.eps}
\hspace{4mm}
\includegraphics[angle=0,width=0.48\linewidth,clip]%
                {pff_em_vs_q2_mud0050_ms0080.phys.eps}

\vspace{-2mm}
\caption{
   Left panel: 
   effective value $\emFpip\!(\Delta t,\Delta t^\prime;q^2)$
   with local (open symbols) and smeared operators for pion source and sink
   (filled symbols).
   Note that we can take arbitrary combinations of $(\Dt,\Dtp)$ 
   by using the all-to-all propagator.
   Right panel :
   $\emFpip\!(q^2)$ as a function of $q^2$.
   We also plot the $q^2$ dependence expected from the VMD hypothesis
   by the dashed line.
}
\label{fig:em:pi:eff+q2-dep}
\end{center}
\vspace{-5mm}
\end{figure}

%// ratio method 

We calculate an effective value of the pion EM form factor from 
\bea
   \emFpip\!(\Delta t,\Delta t^\prime;q^2)
   & = & 
   \frac{2\,M_\pi}{E_\pi(|\bfp|)+E_\pi(|\bfp^\prime|)}
   \frac{R^{\pi J \pi}_{\phi \phi^\prime}(\Dt,\Dtp; q^2)}
        {R^{\pi J \pi}_{\phi \phi^\prime}(\Delta t,\Delta t^\prime; 0)},
   \label{eqn:em:pi:dratio:pff}
   \\
   R^{\pi J \pi}_{\phi \phi^\prime}(\Delta t,\Delta t^\prime; q^2)
   & = &
   \frac{C^{\pi J \pi}_{\phi \phi^\prime}
         (\Delta t,\Delta t^\prime; \bfp,\bfp^\prime)}
        {C^{\pi \pi}_{\phi \phi_l}(\Delta t;\bfp)\,
         C^{\pi \pi}_{\phi_l \phi^\prime}(\Delta t^\prime;\bfp^\prime)},
   \label{eqn:em:pi:ratio}
\eea       
where we use the dispersion relation 
to estimate $E_\pi(|\bfp|)\!=\!\sqrt{M_\pi^2+\bfp^2}$ at $|\bfp|\!\ne\!0$.
As shown in the left panel of Fig.~\ref{fig:em:pi:eff+q2-dep}
we can reliably identify the plateau of $\emFpip\!(\Delta t,\Delta t^\prime;q^2)$
by using different combination of the smearing functions $(\phi,\phi^\prime)$.
The EM form factor $\emFpip\!(q^2)$ is determined by a constant fit 
to $\emFpip\!(\Delta t,\Delta t^\prime;q^2)$.
The statistical accuracy is typically a few \% 
because of the average over the source location mentioned above.
Although we ignore the finite volume correction to $\emFpip\!(q^2)$
in this preliminary report, 
it turned out to be comparable with the statistical error of a few \% level
in our previous study in two-flavor QCD 
at similar values of $m_{ud}$ and $N_s a$ \cite{PFF:Nf2:RG+Ovr:JLQCD}.

%// q^2 dependence 

As seen in the right panel of Fig.~\ref{fig:em:pi:eff+q2-dep},
the $q^2$ dependence of $\emFpip\!(q^2)$ 
is close to the $\rho$ meson pole $1/(1\!-\!q^2/M_\rho^2)$
expected from the vector meson dominance (VMD) hypothesis.
We then assume that the small deviation due to the higher poles or cuts
can be approximated by a polynomial of $q^2$,
and use a parametrization
\bea
   \emFpip\!(q^2)
   & = & 
   \frac{1}{1-q^2/M_{\rho}^2} + c_1\,q^2 + c_2\,(q^2)^2 
   \hspace{1mm} = \hspace{1mm}
   1 + \frac{1}{6} \cradpip q^2 + \cdots.
   \label{eqn:em:pi:q2-dep}
\eea
to determine the pion charge radius $\cradpip$.
Since the deviation of $\emFpip(q^2)$ from VMD is small,
we obtain reasonable $\chi^2/{\rm d.o.f.}\!\sim\!1$, 
and $\cradpip$ does not change significantly by 
the inclusion of higher order corrections of $O(q^6)$.

%// chiral extrapolation

At NLO of $SU(3)$ ChPT, $\cradpip$ is given by \cite{LMFF:ChPT:SU3:NLO}
\bea
   \cradpip
   & = & 
   \frac{1}{2NF_0^2}\left( -3 + 24N\,L_9^r \right)
   -2\nu_\pi -\nu_K,
   \hspace{3mm}
   \nu_P 
   = 
   \frac{1}{2NF_0^2}\ln\left[ \frac{M_P^2}{\mu^2} \right]
   \hspace{3mm}
   (P\!=\!\pi, K),
   \label{eqn:em:pi:chiral_fit:nlo}
\eea
\FIGURE{
   % \centering
   % \begin{figure}[h!]
   % \begin{center}
   % \hspace{2mm}
   \label{fig:em:chiral_fit:pi}
   \includegraphics[angle=0,width=0.48\linewidth,clip]%
                   {r2_em_pff_vs_Mpi2.eps}
   \vspace{-3mm}
   \caption{
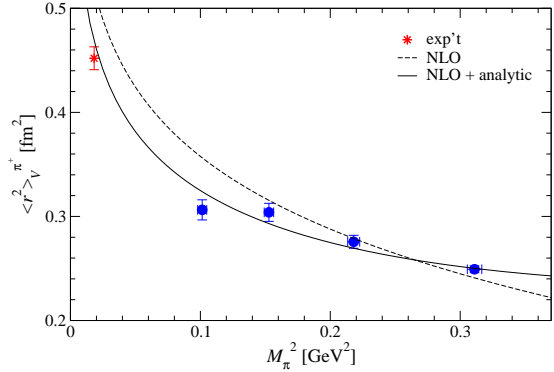
      Chiral extrapolation of $\cradpip$ 
      using the one-loop formula (\protect\ref{eqn:em:pi:chiral_fit:nlo}) 
      (dashed line) 
      and that with a higher order analytic term (solid line).
      The experimental value \cite{PDG} is plotted by a star.
   }
}
\noindent
where $N\!=\!(4\pi)^2$ and the renormalization scale $\mu$ is set to $4\pi F_0$.
We fix $F_0$ to 52~MeV determined from our study of 
the meson decay constants \cite{Spectrum:Nf3:RG+Ovr:JLQCD:lat10}.
This is significantly smaller than the phenomenological value $\sim 88$~MeV
\cite{LECs:Nf3} and enhances the chiral logarithms $\nu_{\pi,K}$.
As shown in Fig.~\ref{fig:em:chiral_fit:pi},
the NLO formula fails to reproduce the quark mass dependence 
of our data (dashed line). 
% with $\chi^2/{\rm d.o.f.}\!\sim\!11$. 
The extrapolation becomes consistent with experiment
% with smaller $\chi^2/{\rm d.o.f}\!\lesssim\!3$
by including a NNLO analytic term $\propto M_\pi^2$
(solid line).
We note that significant NNLO contributions have been observed 
also in our two-flavor studies in a similar region of $m_{ud}$
\cite{PFF:Nf2:RG+Ovr:JLQCD}.

\begin{figure}[b]
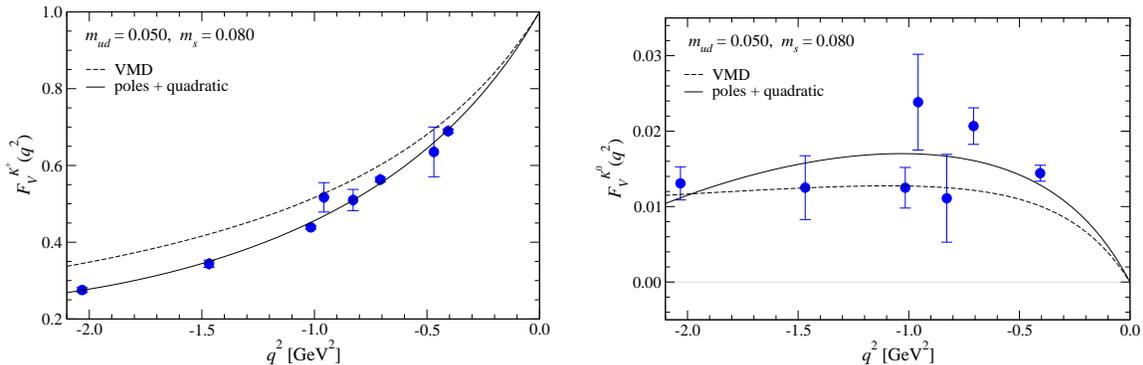

\begin{center}
\includegraphics[angle=0,width=0.48\linewidth,clip]%
                {k+ff_em_vs_q2_mud0050_ms0080.eps}
\hspace{4mm}
\includegraphics[angle=0,width=0.48\linewidth,clip]%
                {k0ff_em_vs_q2_mud0050_ms0080.eps}
\vspace{-8mm}
\caption{
   Electromagnetic form factors of charged (left panel) and neutral
   kaons (right panel) as a function of $q^2$.
   Dashed lines show the pole dependence of 
   Eq.(\protect \ref{eqn:em:k:poles}).
}
\label{fig:em:k:q2-dep}
\vspace{-5mm}
\end{center}
\end{figure}

\begin{figure}[t]
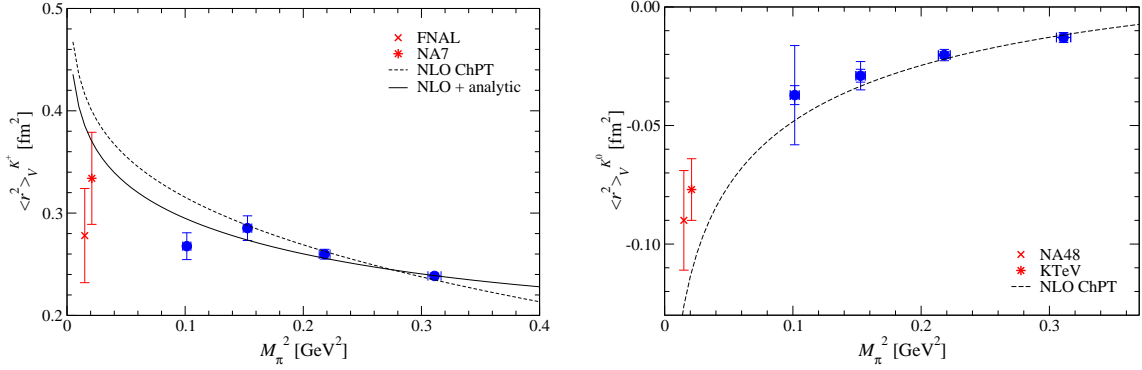

\begin{center}
\includegraphics[angle=0,width=0.48\linewidth,clip]%
                {r2_em_k+ff_vs_Mpi2.eps}
\hspace{4mm}
\includegraphics[angle=0,width=0.48\linewidth,clip]%
                {r2_em_k0ff_vs_Mpi2.eps}
\vspace{-8mm}
\caption{
   Chiral extrapolation of charge radii of 
   charged (left panel) and neutral kaons (right panel).
   Crosses and stars are experimental values \cite{PDG}.
}
\label{fig:em:k:chiral_fit}
\vspace{-5mm}
\end{center}
\end{figure}

%// Kaon EM form factors

The kaon EM form factors are calculated from the ratios 
(\ref{eqn:em:pi:dratio:pff}) and (\ref{eqn:em:pi:ratio})
but with kaon source and sink. 
Results are plotted as a function of $q^2$ in Fig.~\ref{fig:em:k:q2-dep}.
The neutral kaon form factor $\emFkn\!(q^2)$ originates
from a difference between the contributions 
from the down ($\bar{d}\gamma_\mu d$) 
and strange quark currents ($\bar{s}\gamma_\mu s$),
and is much smaller than the charged one $\emFkp\!(q^2)$.
We obtain significant signal for $\emFkn(q^2)$ 
with our statistical accuracy improved by using the all-to-all propagator.
Similar to $\emFpip\!(q^2)$,
the $q^2$ dependence of both $\emFkp\!(q^2)$ and $\emFkn\!(q^2)$ 
is close to that of VMD
\bea
   \emFkp\!(q^2)
   & = &
   \frac{2}{3}\frac{1}{1-q^2/M_\rho^2}
  +\frac{1}{3}\frac{1}{1-q^2/M_\phi^2},
   \hspace{5mm}
   \emFkn\!(q^2)
   =
  -\frac{1}{3}\frac{1}{1-q^2/M_\rho^2}
  +\frac{1}{3}\frac{1}{1-q^2/M_\phi^2}.
   \hspace{3mm}
   \label{eqn:em:k:poles}
\eea
We determine charge radii, $\cradkp$ and $\cradkn$,
using a fitting form with these vector meson poles 
plus a polynomial correction up to $O(q^4)$.

Figure~\ref{fig:em:k:chiral_fit} shows
the chiral extrapolation of $\cradkp$ and $\cradkn$
based on NLO ChPT
\bea
   \cradkp
   & = &
   \frac{1}{2NF_0^2}\left( -3 + 24 N L_9^r \right)
  -\nu_\pi - 2\nu_K,
   \hspace{5mm}
   \cradkn
   = 
   \nu_\pi - \nu_K.
   \label{eqn:em:k:chiral_fit:nlo}
\eea
For $K^+$, 
we again observe that the NLO fit leads to 
a large value of $\chi^2/{\rm d.o.f.}\!\sim\! 3.2$.
The extrapolation becomes closer to experiment 
with acceptable $\chi^2/{\rm d.o.f.}$ ($\sim$\,0.6)
by including a NNLO analytic term. % $\propto M_\pi^2$.

Since $K^0$ does not directly couple to photons, 
the NLO expression of $\cradkn$
does not have analytic terms and hence $O(p^4)$ LECs.
The dashed line in the right panel of Fig.~\ref{fig:em:k:chiral_fit}
is a parameter-free prediction with $F_0$ determined from the decay constants.
Our data are consistent with this NLO prediction. 
For a more rigorous comparison,
calculations with twisted boundary conditions % \cite{TBC}
are currently underway to reduce the large systematic uncertainty 
due to the lack of data near $q^2\!=\!0$.

\vspace{-1mm}
\section{Pion scalar form factor}
\vspace{-2mm}

\begin{figure}[b]
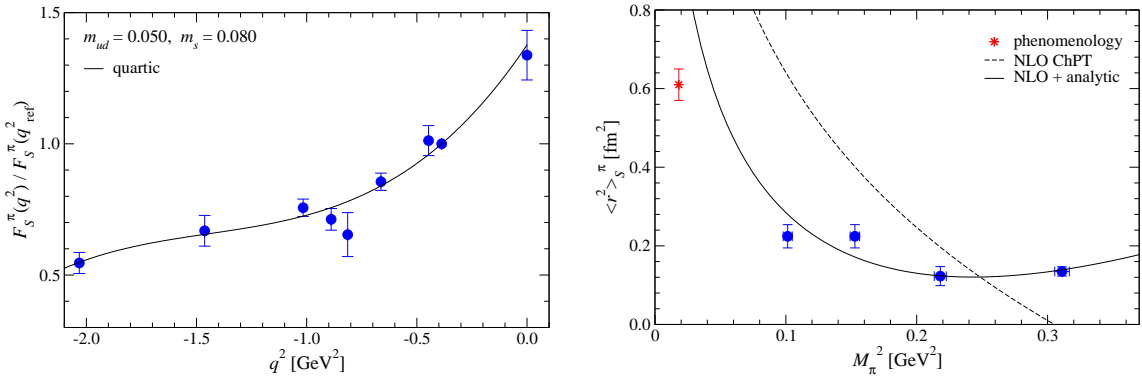

\begin{center}
\includegraphics[angle=0,width=0.48\linewidth,clip]%
                {pff_s_l_vs_q2_mud0050_ms0080.eps}
\hspace{4mm}
\includegraphics[angle=0,width=0.48\linewidth,clip]%
                {r2_s_l_pff_vs_Mpi2.eps}
\vspace{-7mm}
\caption{
   Left panel: 
   $\sFpi(q^2)/\sFpi(q^2_{\rm ref})$ as a function of $q^2$. 
   Right panel: 
   chiral extrapolation of $\sradpi$.
   The star shows a phenomenological estimate 
   from $\pi\pi$ scattering \cite{srad:ChPT:Nf2}.
}
\label{fig:scalar:q2-dep+chiral_fit}
\vspace{-5mm}
\end{center}
\end{figure}

%// scalar form factor 

In this report, we consider 
the scalar form factor normalized at a reference value of $q^2_{\rm ref}$,
namely $\sFpi(q^2)/\sFpi(q^2_{\rm ref})$, 
since it has sufficient information to determine the scalar radius $\sradpi$
and does not need a non-perturbative renormalization of $S$.
We set $|q^2_{\rm ref}|$ to our smallest non-zero $|q^2|$,
where we do not need to subtract the contribution 
from the vacuum expectation value of $S$.
In the left panel of Fig.~\ref{fig:scalar:q2-dep+chiral_fit},
we plot $\sFpi(q^2)/\sFpi(q^2_{\rm ref})$ determined from ratios 
similar to Eqs.~(\ref{eqn:em:pi:dratio:pff}) and (\ref{eqn:em:pi:ratio})
(see Ref.~\cite{PFF:Nf2:RG+Ovr:JLQCD} for details).
Due to the lack of the knowledge of scalar resonances 
at the simulated quark masses, 
we parametrize the $q^2$ dependence of $\sFpi(q^2)$ 
by a generic polynomial form 
\bea
   \sFpi(q^2)
   & = & 
   \sFpi(0)
   \left( 
      1 + \frac{1}{6} \sradpi\, q^2 + d_2 (q^2)^2 + d_3 (q^2)^3 + d_4 (q^2)^4
   \right).
   \label{eqn:scalar:q2-dep}
\eea
We then fit results for $\sradpi$ to the NLO chiral expansion
\bea
   \sradpi
   & = &
   \frac{1}{NF_0^2}
   \left\{
      -8 + 24 N \left( 2\,L_4^r + L_5^r \right) 
   \right\} 
   -12 \nu_\pi - 3 \nu_K.
   \label{eqn:scalar:chiral_fit:nlo}
\eea
The pion-loop logarithm is 6 times larger than that in $\cradpip$
and is further enhanced by our small value of $F_0$.
As shown in the right panel of Fig.\ref{fig:scalar:q2-dep+chiral_fit},
the NLO expression has a strong $m_{ud}$ dependence and 
can not describe our data leading to $\chi^2/{\rm d.o.f.}\sim O(100)$.
This is largely reduced to $\sim 7$
by including a NNLO analytic term $\propto M_\pi^2$
suggesting that
the consistency with $SU(3)$ ChPT should be studied 
by including full NNLO corrections 
as in our previous study in two-flavor QCD \cite{PFF:Nf2:RG+Ovr:JLQCD}.

\FIGURE{
   \centering
   % \begin{figure}[h!]
   % \begin{center}
   \includegraphics[angle=0,width=0.47\linewidth,clip]%
                   {xi_drat3k_k2p_vs_dtsrc_mom0100_smr11.eps}
   \vspace{-3mm}
   \caption{
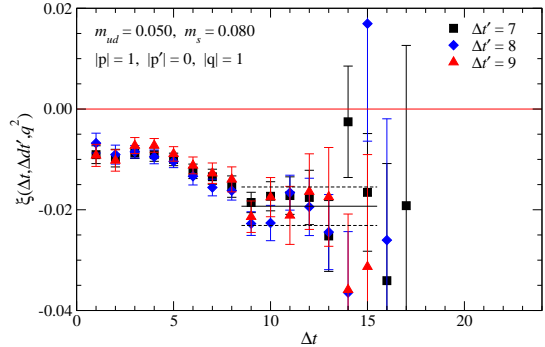
      Effective value of $\xi$ 
      extracted from ratio Eq.~(\protect\ref{eqn:k2pi:drat:3}).
   }
   \label{fig:k2pi:xi}
   % \end{center}
   % \end{figure}
}

\vspace{-1mm}
\section{Kaon weak decay form factors}
\vspace{-2mm}

We calculate the vector and scalar form factors
of the $K \! \to \! \pi$ decays,
namely 
$f_+(q^2)$ and $f_0(q^2)\!=\!f_+(q^2)+f_-(q^2)\, q^2/(M_K^2-M_\pi^2)$,
by using ratios of kaon and pion correlators 
proposed in previous studies \cite{Kl3:drat}.
For instance, 
$\xi(q^2)\!=\!f_-(q^2)/f_+(q^2)$,
which is needed to convert $f_+(q^2)$ to $f_0(q^2)$ (and vice versa),
is determined from a double ratio 
\bea
   % R_k 
   % & = & 
         \frac{ C^{K V_k \pi}_{\phi \phi^\prime}(\Dt,\Dtp,\bfp,\bfpp)
                C^{K V_4 K}_{\phi \phi^\prime}(\Dt,\Dtp,\bfp,\bfpp)   }
              { C^{K V_4 \pi}_{\phi \phi^\prime}(\Dt,\Dtp,\bfp,\bfpp)
                C^{K V_k K}_{\phi \phi^\prime}(\Dt,\Dtp,\bfp,\bfpp)   }.
   \hspace{7mm}
   \label{eqn:k2pi:drat:3}
\eea
This involves the three-point functions with spatial components $V_k$ 
and non-zero meson momenta $\bfp^{(\prime)}$, 
that are quite noisy if naively calculated.
By using the all-to-all propagator, 
we obtain a clear signal for $\xi(q^2)$ as shown in Fig.~\ref{fig:k2pi:xi}.
Both of $f_+(q^2)$ and $f_0(q^2)$ are determined 
with the statistical accuracy of typically a few \% 
even at nonzero $\bfp^{(\prime)}$. %, namely at $q^2 \! < \! (M_K-M_\pi)^2$.

\begin{figure}[t]
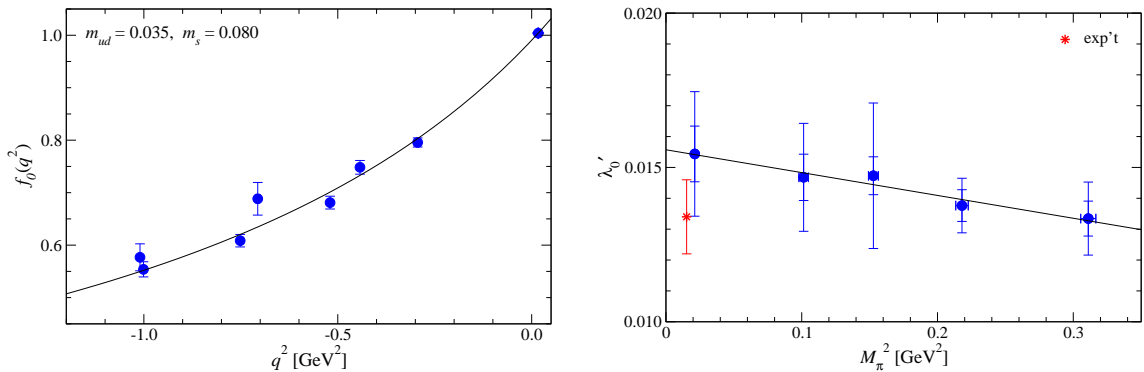

\begin{center}
\includegraphics[angle=0,width=0.48\linewidth,clip]%
                {f0_vs_q2_mud0035_ms0080.phys.eps}
\hspace{4mm}
\includegraphics[angle=0,width=0.48\linewidth,clip]%
                {lambda0_vs_Mpi2.eps}
\vspace{-7mm}
\caption{
   Left panel: $f_0(q^2)$ as a function of $q^2$.
   The solid line shows the quadratic parametrization
   in Eq.~(\protect\ref{eqn:k2pi:q2-dep}).
   Right panel: 
   linear chiral extrapolation of $\lambda_0^\prime$
   compared with experiment \cite{PDG}.
}
\label{fig:chiral_fit:r2_c_v:nnlo}
\vspace{-7mm}
\end{center}
\end{figure}

We plot $f_0(q^2)$ as a function of $q^2$
in the left panel of Fig.\ref{fig:chiral_fit:r2_c_v:nnlo}.
The $q^2$ dependence of both $f_{+,0}(q^2)$ is well 
described by either the single pole or quadratic form
\bea
   f_X(q^2)
   & = &
   \frac{f_X}{1-q^2/M_X^2},
   \hspace{5mm}
   f_X(q^2)
   = 
   1 + c_X  q^2 + d_X (q^2)^2
   \hspace{5mm}
   (X=+,0),
   \label{eqn:k2pi:q2-dep}
\eea   
The normalized slopes $\lambda^\prime_X\!=\!M_{\pi,\rm phys}^2 c_X$ 
are measured in recent experiments \cite{PDG}. 
Our results $\lambda^\prime_{+}\!=\!2.01(25) \times 10^{-2}$
and $\lambda^\prime_{0}\!=\!1.54(20) \times 10^{-2}$ 
extrapolated to $M_{\pi,\rm phys}$ 
are in good agreement with the experiments
as shown in the right panel of Fig.~\ref{fig:chiral_fit:r2_c_v:nnlo}.
We also observe that 
the normalized curvature 
$\lambda^{\prime\prime}_+ =2M_{\pi,\rm phys}^4 d_+ = 0.08(10) \times 10^{-2}$
is also consistent with the experimental value $0.20(5) \! \times \! 10^{-2}$.

%// conclusion: ===============================================================

\vspace{-1mm}
\section{Conclusions} % 0.5p
\vspace{-2mm}

We report on our calculation of the light meson form factors
in three-flavor QCD with overlap quarks.
For the EM and scalar form factors,
we observe that the mild $m_{ud}$ dependence of our data
can not be described by NLO ChPT.
We are planning to extend our analysis to NNLO.
To this end, 
it is helpful to calculate various observables
in order to constrain many $O(p^4)$ and $O(p^6)$ LECs 
involved in NNLO chiral expansions. % \cite{LMFF:ChPT:SU3:NNLO}.
For instance, we calculate the kaon EM form factors
with small additional costs using the all-to-all propagator.

We also confirm that the shape of the $K \! \to \! \pi$ decay 
form factor is in good agreement with experiment. 
Our calculations are being extended to a larger volume 
with twisted boundary conditions 
in order to carry out a chiral extrapolation of $f_+(0)$ 
with controlled systematic uncertainties,
which is essential for a reliable estimate of $|V_{us}|$.

%// Acknowledgments -----------------------------------------------------------

\vspace{3mm}

Numerical simulations are performed on Hitachi SR11000 and 
IBM System Blue Gene Solution 
at High Energy Accelerator Research Organization (KEK) 
under a support of its Large Scale Simulation Program (No.~09/10-09).
This work is supported in part by the Grant-in-Aid of the
Ministry of Education, Culture, Sports, Science and Technology
% (No.~20105001, 20105002, 20105003 20105005, 21674002, 21684013 and 220340047).
(No.~21674002, 21684013 and 220340047)
and the Grant-in-Aid for Scientific Research on Innovative Areas
(No.~20105001, 20105002, 20105003 and 20105005)
\vspace{-1mm}

%// references: ===============================================================

\end{document}